\documentclass[10pt,conference,a4paper]{IEEEtran}
\usepackage{natbib}
\usepackage{times,amsmath,epsfig}
\title{A new approach for Trading based on Long-Short Term memory technique}
\author{%
{Zineb Lanbouri{\small $~^{\#1}$}, Saaid Achchab{\small $~^{*2}$} }%
\vspace{1.6mm}\\
\fontsize{10}{10}\selectfont\itshape
$^{\#}$\,National School for Computer Science and Systems analysis,\\ Mohamed V University,\\ Rabat, Morocco\\
\fontsize{9}{9}\selectfont\ttfamily\upshape
$^{1}$\,lanbourizineb05@gmail.com\\
$^{2}$\,achchab@ensias.ma%
\vspace{1.2mm}\\
\fontsize{10}{10}\selectfont\rmfamily\itshape
\fontsize{9}{9}\selectfont\ttfamily\upshape
}
\begin{document}
\maketitle
\begin{abstract} 
Stock market prediction has always been crucial for stakeholders, traders and investors. We developed an ensemble Long Short Term Memory (LSTM) model that includes two time frequencies (annual and daily parameters) in order to predict next day Closing price (one step ahead). Based on a four-step approach, this methodology is a serial combination of two LSTM algorithms.
The empirical experiment is applied to 417 NY stock exchange companies. Based on Open High Low Close metrics and other financial ratios , this approach prooves that the stock market prediction can be improved.
\end{abstract}

%
\section{Introduction}
\begin{figure}[htbp]
\centerline{\includegraphics[scale=0.7]{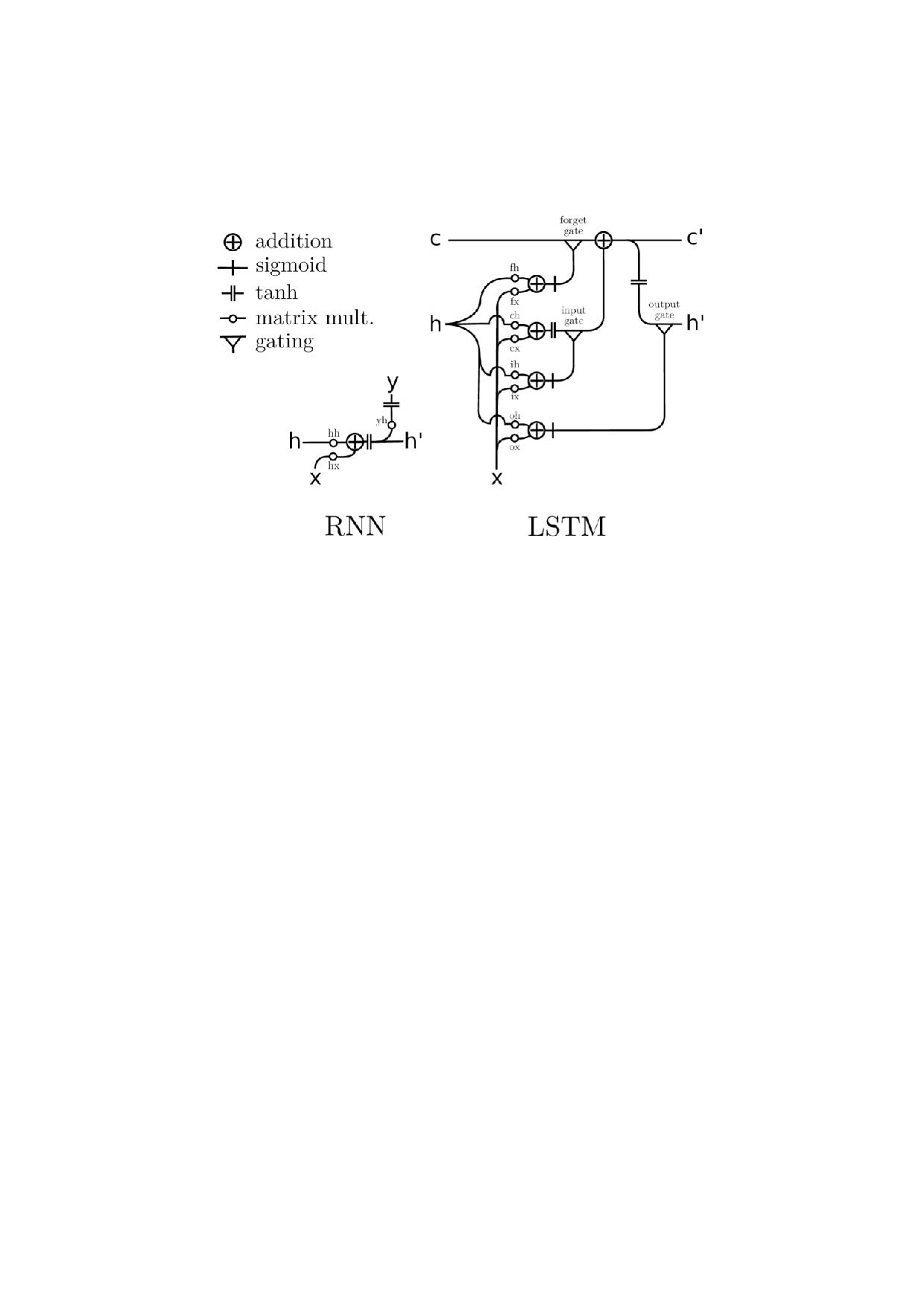}}
\caption{LSTM as an improvment of RNN}
\label{lstm}
\end{figure}

Financial Times Series Forecasting is a crucial topic in finance and its prediction is an extensive ongoing research issue where computer science algorithms are very relevant. "It was estimated that, in 2012, approximately 85\% of trades within the US Stock markets were performed by algorithms" \cite{Glantz13}.
To date, there has been numerous works linking Machine Learning to financial decisions and specially to trading strategies. \cite{Dixon16} shows that "Back Propagation and Gradient Descent have been the preferred method for training finance structures due to the ease of implementation and their tendancy to converge" \cite{Fehrer15, Teixeira10}.  \\
\cite{Kodia10} focuses on social interactions using a multi-agent based simulation. \cite{Batres15}'s paper introduces financial time series in a comparison between Logistic regression, MLP and Naive Bayes.\\
Deep Learning outperforms in finance as well, and Long-Short Term Memory (LSTM) is very promising in forecasting field because  of its ability to memorize data.
"Capturing spatio-temporal dependencies, based on regularities in the observations, is therefore viewed as a fundamental goal for Deep Learning systems." \cite{Arel10}
\\We aim to use a Long Short Term Memory ensemble method with two input sequences, a sequence of daily features and a second sequence of annual features, in order to predict the next day closing price and make a better decision in trading. 
LSTM is the most convenient for the following reasons:
\begin{itemize}
\item LSTM is considered as an improvment of Reccurrent Neural Network which comes as a solution to vanishing and exploding gradient, see figure (Fig ~\ref{lstm})  \cite{Bao17, Werbos88, Schmidhuber97};
\item LSTM is suitable for sequences \cite{Sutskever14};
\item LSTM can store and retrieve information using its gates \cite{Bengio09};
\item LSTM doesn't flow in a single way (unlike Neural Networks);
\item LSTM technique can distinguish between recent and early examples \cite {Nelson17}.
\end{itemize}

New studies focuses on Ensemble methods where the weakness of a method is balanced out by the strength of another to produce high quality \cite{Chang09}. Ensemble learning can be either parallel or serial. A parallel ensemble results from different learners which are combined according to the schemes of Majority Voting, Weighted Majority Voting, Min, Max, etc.. Serial learning arranges different base learners in sequence and selects the result of one learner as the final output. Our study relies on heteregenous features with two temporal frequencies, it is then more appropriate to use two LSTM learners to form one ensemble method.
\\The main contributions of this study are the following: \\(1) a new ensemble method where two time frequency inputs can be combined to predict the next day closing stock price; (2) evaluation of the method by comparing it to other Machine Learning techniques and financial strategies. 
\\The paper is divided into four sections. In section II we present some of the related work to this subject that can be found in the literature. Section III details the research methodology, describes the data pre-processing and develops the experiment steps. Section IV analyses the results. Finally, in section V, we draw some conclusions and future work.
The objective of this project is to study the effectivness of an LSTM ensemble method on the stock market prediction based on Open, High, Low, Close and Volume indicators as well as some financial metrics used by traders like MACD and RSI and other financial ratios, evaluate its performance in terms of RMSE and other measures through experiments on real NYSE Stock Exchange data and analyze and compare our results to other machine learning techniques.

\section{Litterature Review} \label{sec:LitteratureReview}

Times series prediction is a field that has been related for years by statistics and Machine Learning techniques.
\subsection{Pure single learner for Stock market prediction}
\subsubsection{Autoregressive integrated moving average (ARIMA)}
Researchers have exploited statistical techniques to predict and forecast financial times series like ARIMA which requires strict assumption like stationarity while finance forecasting is complex, noisy and non-stationary \cite{Bao17}.
\subsubsection{Artificial Neural Network}
 \cite{Kuremoto14}'s paper shows that Artificial Neural Network is used for Times Series Forecasting since 1980s, more than 5000 publications has been released on ANN for Forecasting \cite{Crone07}.
However, ANN presents issues like overfitting, impact of initial values and $alpha$.
\\In order to tackle these issues, Deep Neural Network could be considered as a solution.
\subsubsection{Deep Neural Network}
DNN is effectively more robust toward overfitting and can model complex non linear relationship between dependant and independant variables \cite{Arel10}. Howerver, \cite{Schmidhuber15} explains some of the drawbacks of Deep Neural Network as: \\-Slow Convergence time; \\-Vanishing or exploding gradient; \\-Expensive computation.
\subsubsection{Deep Learning}
\cite{Hinton06} used a Deep Belief Network and gets 1,25\% test error. Deep Learning has the following advantages:\\ -Ability to learn complexity;\\-Aptitude to learn with little human input with low level/intermediate/high level of abstraction;\\-Strong unsupervised learning.

\subsection{Ensemble and hybrid learning models for Stock market prediction}
\cite{Zhang03} apply a hybrid model based on ARIMA and Neural Network to raise the prediction precision, \cite{Werbos74} shows that ARIMA-ANN model outperfoms statistical methods. \\ \cite{Ding15} chooses to combine Neural tensor and Convolutional Neural Network to predict short and long term influences.  \cite{Bao17} use Stacked Auto-encoders and Long Short Term memory.


\section{Methodology}
Our model combines two feature categories in order to transmit the different aspects of an equity to the network. \\The first aspect considers historical daily stock price that reflects its trend. OHLC (Open, High, Low and Close stock prices), volume and trading indicators MACD, RSI and Signal show a share evolution based mainly on supply and demand. \\The second aspect consists of financial metrics that are part of the financial analysis. The objective is to take advantage of both technical analysis and fundamental-financial analysis.
We adopt an Ensemble LSTM approach as explained in the figure below (Fig ~\ref{Ensemble LSTM approach}):
\begin{figure*}[htbp]
\centerline{\includegraphics[scale=0.7]{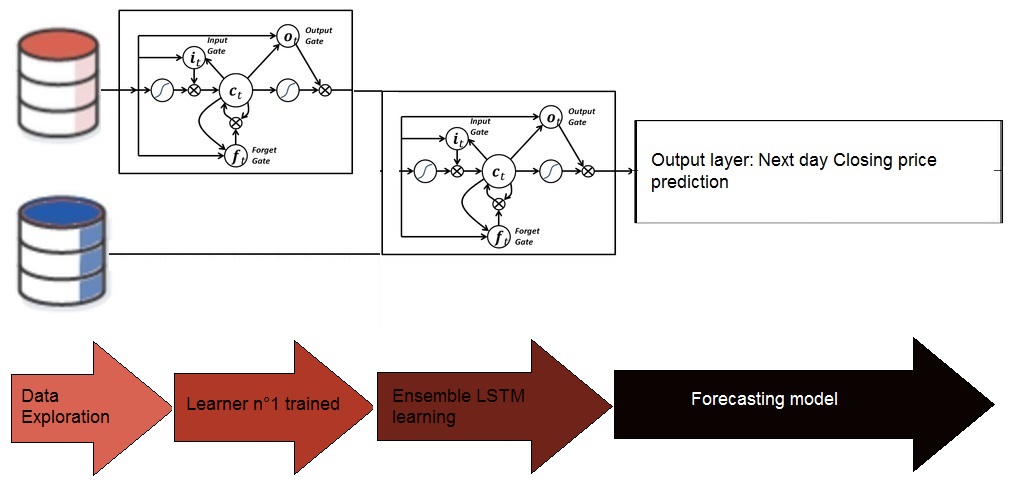}}
\caption{Ensemble LSTM approach}
\label{Ensemble LSTM approach}
\end{figure*}
\subsection{Data exploration}
The first step in this research is Data exploration.
The database used in this research contains New York Stock Exchange S\&P500 data, publicly available at https://www.kaggle.com/dgawlik/nyse, for the period between 04/01/2010 and 30/12/2016. It is splitted into 2 Datasets, one for prices : Open - Low - High - Close - Volume and daily indicators : MACD - Signal - RSI. \\Moving Average Convergence Divergence (MACD) is a trend-following momentum indicator and when MACD crosses its signal, it can function as a buy and sell signal.
Relative Strength indicator (RSI) indicates the internal strength of an equity, it also reflects the vitality of increases over decreases.
\\The second Dataset contains 75 financial annual ratios that shows informations about the company itself. The following ratios are some of them:
\begin{itemize}
\item EPS (Earning Per Share)
\item ROE (Return On Equity)
\item Payout ratio
\item Dividend yield
\item PER (Price Earning Ratio)
\item PBR (Price to Book ratio)
\end{itemize}

Data collected contains 417 companies, the training set consists of consecutive observations from 04/01/2010 to 07/08/2015 while the test set consists of observations from 08/08/2015 to 31/12/2016 which represents sequences of 1408 time series for the training set and 352 for the test set. Before using Data as input for our Ensemble LSTM method, missing values are handled using Mean and Logistic regression functions in Matlab, then z-score normalization is applied (see equation ~\ref{Zscore}).

\begin{equation}
   z =\frac{ (x - \mu)}{\sigma}
\label{Zscore}
\end{equation}

MACD,the Signal and RSI are defined by the following formulas (~\ref{macd}) and (~\ref{rsi}) :
\begin{equation}
\left \{
\begin{array}{r c l}
MACD = \underset{12}{EMA} - \underset{26}{EMA}\\
EMA = (\underset{today}{Value} * \frac{smoothing}{(1+days} ) + (\underset{yesterday}{Value} * (1 - \frac{smoothing}{(1+days} ) )\\
smoothing = \frac{2}{(selected time period + 1)}
\end{array}
\right .
\label{macd}
\end{equation}

\begin{equation}
\begin{array}{r c l}
RSI(step one) = 100 - \frac{100}{1 + \frac{Average Gain}{Average Loss}}\\
RSI = 100 - \frac{100}{1 + \frac{Previous Average Gain * 13 + Current Gain}{Previous Average Loss * 13 + Current Loss}}
\end{array}
\label{rsi}
\end{equation}


\subsection{Hierarchical LSTM technique}
\begin{figure*}[htbp]
\centerline{\includegraphics{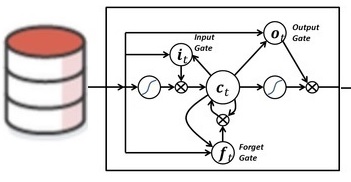}}
\caption{LSTM number 1}
\label{learner1}
\end{figure*}

The chosen Hierachical LSTM technique is based on the assumption that the use of Ensemble method is a way such that the weakness of a method will be balanced out by the strength of another. The second step in our model is to construct the first LSTM learner that can also be considered as the first weak learner (see fig ~\ref{Ensemble LSTM approach}).\\
As a reminder, Long Short Term Memory (LSTM) is a special type of Recurrent Neural Network (RNN) with the capability of learning long term and short term despendencies. Relying on gates, this Deep Learning algorithm has the ability to memorize. It is then well suited for times series prediction (see fig ~\ref{lstm}) and solve the problem of Vanishing gradient.\\
This model contains an LSTM algorithm modeling prediction based on annual features (financial ratios) (see fig ~\ref{learner1}) for the sequence between 2012 and 2016. The input of the first learner consists on 75 features over a 5-sequence (2012 - 2013 -2014 - 2015 - 2016). The output result is a prediction for the closing price that will be considered as one of the inputs of the second LSTM learner.\\
Then, the serial ensemble learning is the result of the combination of the first learner's output (prediction) and the other daily features as an input for the second learner as follows: 
\begin{itemize}
\item Open
\item Highest price for the day
\item Lowest price for the day
\item Close
\item Volume
\item MACD
\item Signal
\item RSI
\item the first learner's prediction
\end{itemize}
The training is then executed over two stages (sequentially).

After training, the model is capable of forecasting and predicting the Closing price one step ahead (see fig ~\ref{Ensemble LSTM approach}).

\section{Results and evaluation}
Experiments were carried out to predict the next day Closing price following the approach described above.
As explained, the Ensemble LSTM approach is composed of one main dataset splitted into two datasets sequentially different since one is for annual features and the other for daily features.\\
The empirical experiment is applied on 417 New York Stock Exchange companies using financial ratios for the first learner. Then the predicted value is included into the input of the second LSTM learner.
This article aims to analyze the hierachical LSTM technique in terms of Root Mean Square Error (RMSE) and Forecasting. 
\subsection{Results}
\begin{figure*}[htbp]
\centerline{\includegraphics{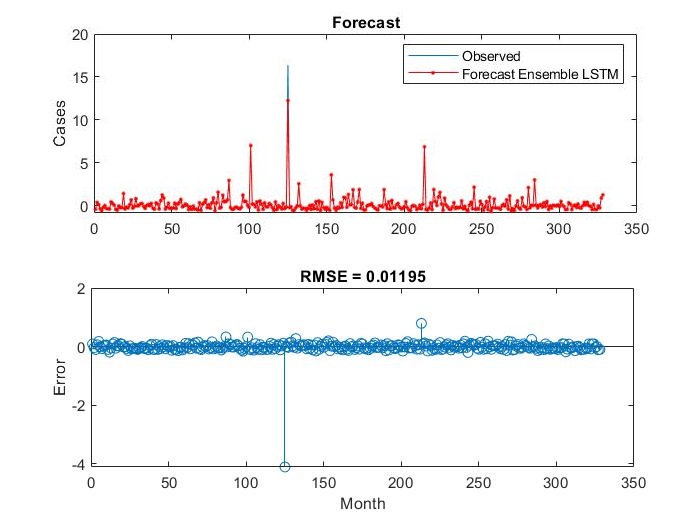}}
\caption{Forecast vs real values when Ensemble LSTM approach is used}
\label{Forecast2}
\end{figure*}

The Ensemble LSTM to be used is a Regression algorithm that allows to predict the next day Closing price. To fix the architecture of our method, several experiences were done to choose the number of hidden units and the learning rate in order to get the best results. The topology chosen is 20 hidden nodes for the first LSTM and an initial learning rate of 0.005, the output is then included into the input of the second LSTM, with a total of 9 features and 200 hidden nodes.\\
Figures ~\ref{Forecast2} presents the Ensemble LSTM model performance in terms of Mean Squared Error and Forecasting. Our model shows a great performance since RMSE is equal to 0.0119 which is more interesting than LSTM standalone for the first Databse alone with 0.0124.

\begin{figure*}[htbp]
\centerline{\includegraphics{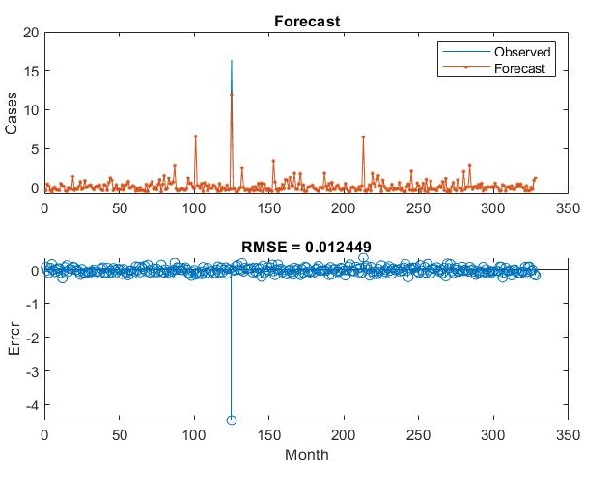}}
\caption{Forecast vs real values when LSTM is applied to DB number 1 alone}
\label{Forecast1}
\end{figure*}

\subsection{Analysis}
\begin{figure*}[htbp]
\centerline{\includegraphics{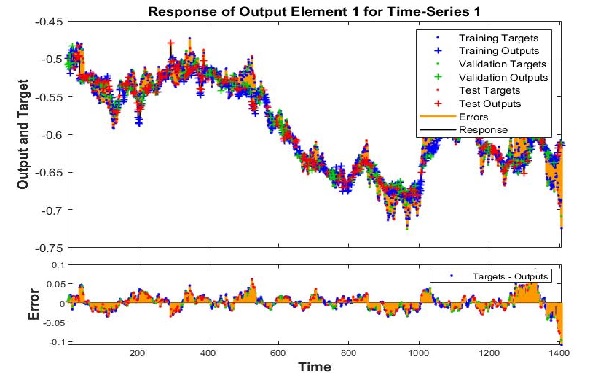}}
\caption{Neural Network response}
\label{nn}
\end{figure*}

In order to determine if there is an improvement while applying an ensemble model, we executed the LSTM algorithm to the first Database (with daily features) based on a sequence to determine the last Clossing price, we then got an RMSE equal 0.0124. Then we applied LSTM standalone to the annual Database and got RMSE equals to 0.08. And finally, to get sure that LSTM is more appropriate in case of Times series, we chose to compare it to Neural Network. We used Neural Network for times series with Matlab, Levenberg-Marquardt as a training algorithm and 10 hidden nodes. We got RMSE equals to 0.067 which is better than LSTM applied to the second Database ~\ref{nn}.
Table ~\ref{rmse} compare our approach to LSTM standalone, using each dataset apart and then to Neural Network.

\begin{table}[h]
\caption{RMSE for the LSTM standalone, LSTM ensemble and Neural Network}
\label{rmse}
\begin{tabular}
{|c|c|c|c|}
\hline
LSTM 1 & LSTM 2 & Ensemble LSTM & NN \\
\hline
0.0124 & 0.08 & 0.0119 & 0.07\\
\hline

\end{tabular}
\end{table}

LSTM's ability to store historic data for a long period and retrieve valuable data when required makes this technique a good technique for time series forecasting.

\section{Conclusion and future work}
In this study, we established an ensemble method based on LSTM algorithm to predict the next day Closing price. This model takes into account two frequencies with two types of variables, daily and annual variables. It confirms then our hypothesis that the combination of these two databases gives a better performance. This ensemble method can be a useful tool for traders and stakeholders to determine their trading strategy.
For future work, we attend to achieve a benchmark between this hierarchical technique and traditional finance methods.
Furthermore, Deep Learning and LSTM are very promising, we can then expect to improve this model and achieve online prediction for forecasting.


\end{document}